# Measures of Basicity in Silicate Minerals Revealed by Alkali Exchange Energetics

**James R. Rustad**

**Corning Incorporated, Corning, NY 14830**

**(Jan 8, 2015)**

**Abstract-** The energies of the alkali exchange reactions ($K_2O$ + Na mineral = $Na_2O$ + K mineral) are computed from density functional calculations for $(Na,K)_4SiO_4$, $(Na,K)_2SiO_3$, $(Na,K)_2Si_2O_5$, $(Na,K)AlSi_3O_8$, and $(Na,K)AlSiO_4$. First-principles calculations are compared against (1) experiment; (2) thermodynamic models; (3) calculations using empirical interatomic potentials (with both polarizable shell model potentials and pairwise potentials with partial charges); and (4) empirical correlations based on optical basicity and Pauling bond strength. The first-principles calculations correlate well with experimental values. The shell model potentials appear to account for Si-O-Si vs. Si-O oxygen basicity, but don't recover the differences in basicity between Al-O-Si and Si-O-Si donors. The pair potentials don't even qualitatively reproduce the reactivity trends established in the first-principles calculations. Empirical correlations based on optical basicity and Pauling bond strength successfully rank-order the exchange energies, but fail to describe the large difference in basicity between Si-O-Si and Si-O oxygen atoms. Knowing which models are capable of predicting ion exchange energies in silicate minerals improves understanding of the physics of bonding in ion-exchanged glasses.

## Introduction

The chemical strengthening of glass results from post-glass production exchange of a large ion (e.g. K) for a small ion (e.g. Na) within the glass network[1]. This exchange can be carried out by placing a sheet of glass in a concentrated molten salt of the larger ion, resulting in a diffusion profile of K/(Na+K) as a function of depth into the glass sheet. The process must take place at a sufficiently low temperature to inhibit relaxation to the equilibrium volume appropriate to the changing K/(Na+K). If the volume cannot fully relax as the potassium replaces the sodium in the glass, pressure will build up along the exchange profile. Strengthening apparently results because this pressure works against the propagation of incipient fractures nucleating at the surface, strengthening the glass under tensional stresses, against which it would otherwise be very weak.

An important factor in the energy balance associated with ion exchange is the enthalpy of the reaction:

$$2KNO_3(s) + Na_2O(gl) = 2NaNO_3(s) + K_2O(gl) \qquad (1)$$

where the notation $Na_2O(gl)$ refers to the sodium-bearing glass and $K_2O(gl)$ represents the exchanged potassium-bearing glass. A little thought reveals that Reaction (1) will be unfavorable, as an oxide glass will almost always be a stronger base than the nitrate and will prefer to bind the smaller sodium ion. The associated free energy, along with the free energy required to build up compressive stress in the glass, will be balanced against the favorable free energy of solution of the $NaNO_3$ in a sufficiently dilute salt bath driven by the entropy of mixing.

Reaction (1) can be expressed as the sum two reactions:

$$2KNO_3(s) + Na_2O(s) = 2NaNO_3(s) + K_2O(s) \qquad (2)$$

$$K_2O(s) + Na_2O(gl) = Na_2O(s) + K_2O(gl) \qquad (3)$$

Reaction (2) is a simple ligand exchange reaction. The reaction is strongly endothermic because the smaller sodium ion will out-compete the larger potassium ion for the more basic oxide ion, leaving potassium to bind with the more acidic nitrate ion. The Materials Genome Project database[2] gives a value of +107 kJ/mol (measured from calorimetry) for Reaction 2. Reaction (3) is expected to be exothermic, because the glass will almost always be a weaker base than the oxide ion. The reason for decomposing Reaction (1) into the sum of Reaction (2) and Reaction (3) is that if the enthalpy of Reaction (2) is known, all attention can be focused on Reaction (3). From a modeling perspective, Reaction (3) is simpler because a representation of the nitrate is not needed; in essence, the reference compound has been shifted from nitrate to oxide. There is no loss of generality as the enthalpy of Reaction (2) is available over a range of temperature.

Quantitative prediction of the enthalpy of Reaction 3 as a function of glass composition requires complex thermodynamic models that do not yet exist for the range of compositions of typical ion exchange glasses. However, it is clear that Reaction (3) would be made less exothermic by increasing the basicity of the glass to make the binding in the glass matrix more competitive with the oxide ion. To illustrate, consider the orthosilicate composition $Na_4SiO_4$. Since the $SiO_4^{-4}$ silicate anion is nearly as good a base as the oxide ion, Reaction (3) will not be so favorable (i.e. the preference of the sodium ion for the oxide ion over the orthosilicate ion will not be very large). On the other hand, for the tectosilicate nepheline composition, $NaAlSiO_4$, the Al-O-Si bridging oxygen atoms are relatively poor bases and the enthalpy of Reaction 3 will be much more favorable (i.e. the smaller sodium will strongly prefer the oxide ion to the Al-O-Si

"ligands" in the glass).   Likewise, the exchange reaction involving $NaAlSi_3O_8$ should be even more negative as the Si-O-Si linkages (making up 4/9 of the bonds into the alkali ions) will be poorer bases than the Al-O-Si linkages (making up 5/9 of the bonds into the alkali ions).

In the absence of thermodynamic models, one could look to atomistic models for making predictions of Reaction (3) as a function of melt composition.  One could make an atomistic model of a glass, and a molten salt, and, from these, estimate exchange energies as a function of glass composition.  How could atomistic models be tested and validated?  There is no reason we have to use glasses in Reaction 3; solid phases will do as well to illustrate the reactivity trends.  In contrast to the glasses, these enthalpies are available from calorimetry and any proposed model for ion exchange in the glass had better be able to accurately calculate them.  In this paper, ion exchange enthalpies are computed for $Na(K)_4SiO_4$, $Na(K)_2SiO_4$, $Na(K)_2Si_2O_5$, $Na(K)AlSiO_4$, $Na(K)AlSi_3O_8$ (structures are shown in Figure 1) using (1) the FactSage 6.4 thermodynamic modeling program (Thermfact Ltd. Montreal, Canada), (2) first principles methods including both Materials Genome Project (MGP) database[2] and CASTEP[3] as implemented in Materials Studio) and (3) empirical potentials using the GULP code[4] in conjunction with a rigid-ion pair potential from Pedone and co-workers[5] (PMMCS) and polarizable shell-model potentials from the Catlow group library[6].

**Results**

The calculated values of $\Delta E$ (electronic energy difference at 0K) for the five alkali exchange reactions are given in Table 1.  For comparison with the MGP, CASTEP is used to compute the reaction energies for all the compounds in Table 1 (taking the optimized structures in the MGP as the starting values for the full (lattice parameters + coordinates) geometry optimization in CASTEP (Perdew-Burke-Ernzerhof exchange-correlation functional[7], ultrasoft pseudopotentials[8], cutoff energy = 370 eV).  As shown in Table 1, the level of agreement between CASTEP and MGP is good; the main difference probably is probably due the projector-augmented-wave[9] (PAW) technique used in the MGP versus the ultrasoft pseudopotentials used in CASTEP.  This comparison gives useful information about what kinds of variations might be expected between slightly different density functional theory (DFT) electronic structure calculations using the same exchange-correlation functional.

Experimental values are provided from the Materials Genome Project Database and the NIST/NBS tables.[10]  It should be noted that the Na silicate and K silicate phases for a given alkali/silicon ratio, are, in general, *not* isostructural; the aim here is not to mimic the process of ion exchange, but to evaluate which methods are capable of reproducing the experimentally measured reactivity trends for ion exchange between the crystalline compounds.

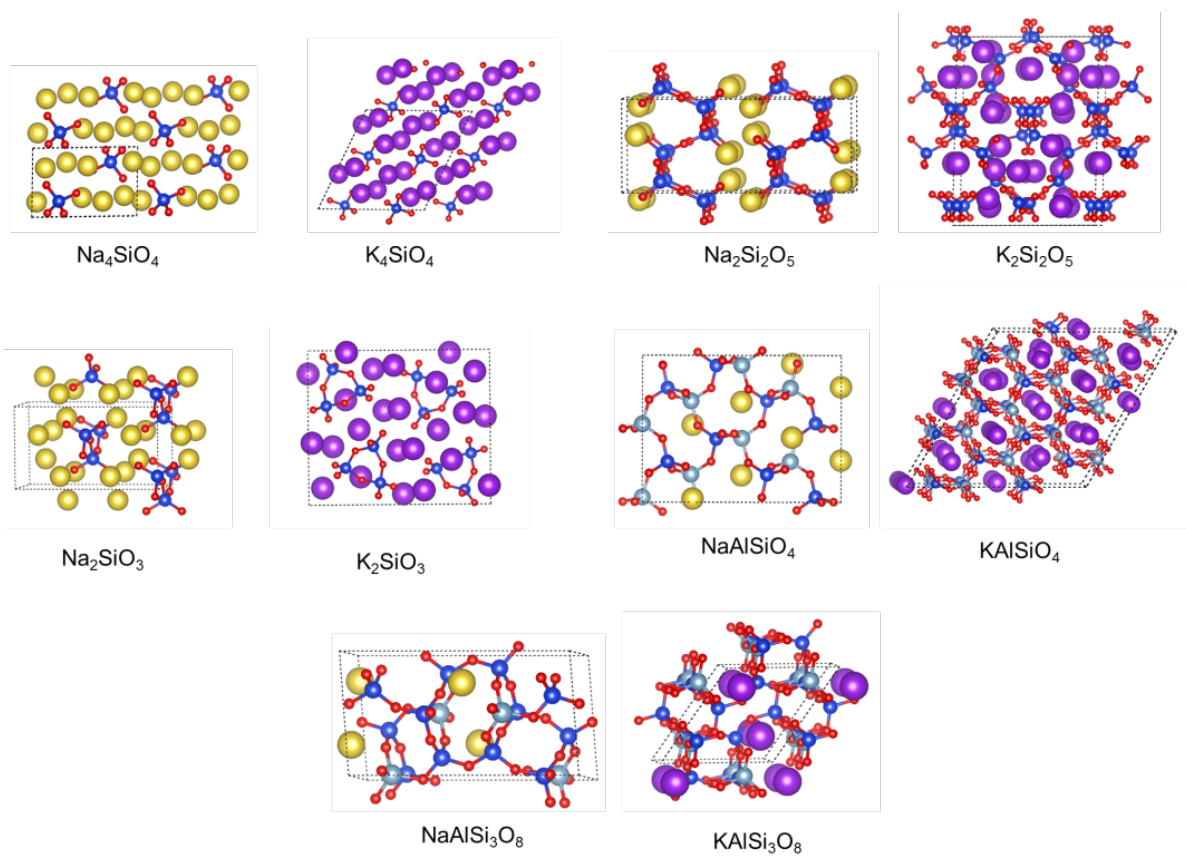

**Figure 1.** Phase pairs considered for calculating ion exchange enthalpies in this paper. Large yellow spheres are sodium, large purple spheres are potassium. The aluminosilicate framework is shown with a ball-and-stick model with oxygen represented with small red spheres, silicon as small dark blue spheres, and aluminum as small light blue spheres. Structures taken from the Materials Genome Project database.

The first-principles calculations exhibit a clear decrease in the favorability of the exchange reaction as the number of non-bridging oxygen atoms increases, (ΔE $Na_4SiO_4$ > (less negative than) ΔE $Na_2SiO_3$ > ΔE $Na_2Si_2O_5$). This is expected: the non-bridging oxygen atoms make stronger bases than the bridging oxygen atoms, therefore, the higher the fraction of non-bridging oxygen atoms, the less negative the reaction enthalpy. In other words, the non-bridging Si-O ligands more closely approach the oxide ion in terms of $M^+$-O bond strength, and therefore have a higher affinity for the smaller $Na^+$ cation (if they were just as good as the oxide ion, the change in energy would be zero). For the tectosilicate feldspar-type ($MAlSi_3O_8$) and carnegieite/kaliophilite-type ($MAlSiO_4$) compositions, the calculations indicate that the feldspar composition (4/9 Si-O-Si; 5/9 Al-O-Si) has a more negative enthalpy than carnegieite composition (all Al-O-Si bridges).

Table 1. $\Delta H$(298.15 K) (kJ/mol) (or $\Delta E$ at 0K for atomistic models) for alkali exchange reactions for different mineral phases. Energies are for one mol of alkali.

| | FS | MGP | CASTEP[a] | PMMCS | SHELL | exp |
|---|---|---|---|---|---|---|
| $K_2O + 2NaNO_3 = Na_2O + 2KNO_3$ | +108 | +112 | | | | +107 |
| $K_2O + 0.5Na_4SiO_4 = Na_2O + 0.5K_4SiO_4$ | | -26 | -25 | -19 | -43 | |
| $K_2O + Na_2SiO_3 = Na_2O + K_2SiO_3$ | -55 | -41 | -40 | 11 | -55 | -82[b] |
| $K_2O + Na_2Si_2O_5 = Na_2O + K_2Si_2O_5$ | -107 | -56 | -53 | -5 | -71 | -90[b] |
| $K_2O + 2NaAlSiO_4$ c$= Na_2O + 2KAlSiO_4$[d] | -95 | -109 | -98 | -46 | -138 | -110[e] |
| $K_2O + 2NaAlSi_3O_8$[f] $= Na_2O + 2KAlSi_3O_8$[g] | -102 | -138 | -133 | 18 | -136 | -119[e] |

[a]values in parenthesis calculated using CASTEP (ultrasoft PBE pseudopotentials $E_{cut}$=370 eV); [b]values from MGP database; [c]carnegeite; [d]kaliophilite (hexagonal); [e]values from Ref 10; [f]low albite; [g]microcline

**Discussion**

The $\Delta E$ values from first-principles calculations and the $\Delta H$ (298.15 K) experimental values are in qualitative agreement with the ordering expected on intuitive ideas about the basicity of the solid phases. Although (for the four points available) the first-principles calculations *correlate* well with the experimental values in the MGP and in Ref. 10 (see Figure 2), the slope is far from unity (~2.6 kJ/mol) and the intercept is far from zero (+177.8 kJ/mol).

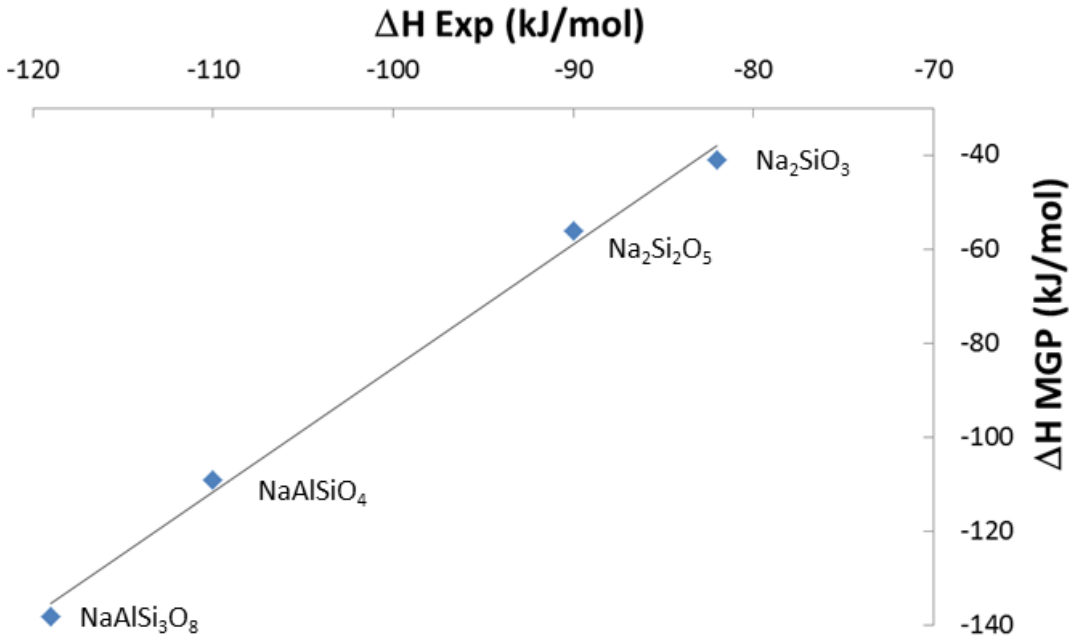

Figure 2. Correlation between first-principles calculations in the Materials Genome Project database and measured enthalpies of the alkali exchange reactions for various phases

The FactSage calculations for ΔH (298.15) track fairly well with the experimental values, except for the very large negative enthalpy of the reaction for $Na_2Si_2O_5$. The PMMCS force field fails to even qualitatively reproduce the trends established from the first-principles calculations. The fact that one can *guess* the exchange energy versus basicity trend better than it can be calculated by the PMMCS force field is stunning. One would expect the qualitative ideas about the relative bonding strength of bridging versus non-bridging oxide ions and Al-O-Si versus Si-O-Si bridges should be qualitatively borne out in the empirical potential calculations as one's intuition is essentially driven by size/charge ratios which should come out reasonably well in the pair potentials. The structures are reasonably well-predicted[5]. The problem isn't just referencing all calculations to the pure alkali oxide (which is a more difficult phase to treat for the ionic model), as this would just shift the entire correlation up or down. The correlation would not improve if all calculations were referenced to $(Na,K)_2SiO_3$ instead of $(Na,K)_2O$. The polarizable shell model potentials seem to do much better (see Figure 3), except that, surprisingly, the $NaAlSi_3O_8$ exchange reaction is less negative than $NaAlSiO_4$. The failure of both potential models to be able to distinguish the higher basicity of the carnegeite/kaliophilite composition relative to the feldspar composition is disconcerting in that, again, one feels that one can guess the sense of the reaction better than it can be calculated with the empirical atomistic potentials.

If intuitive guesses seem as good, or better than, estimates from the empirical potentials, a possible approach is to make more quantitative estimates of melt basicity using empirical models. Table 2 compares the results from the first principles calculations in the MGP database and (1) optical basicity[11] and (2) average Pauling bond strength (PBS) of the oxide ion (excluding contributions from the alkalis). Optical basicity is determined spectroscopically as discussed in Ref. 11. PBS is the sum, over all cations bound to a given oxygen atom, of the charge/coordination number of each bound cation (not including alkalis). The results are shown Figure 4.

While the empirical types of approaches can be used to rank-order the materials, both suffer from the small difference in both optical basicity and PBS between $Na_2Si_2O_5$ and $NaAlSiO_4$, contrasted with the large difference in the alkali exchange energy calculated for these phases with the first principles calculations. This failure likely results from the very different behavior of bridging and non-bridging oxygen atoms in terms of the alkali exchange thermodynamics.

It could be argued, from a Pauling bond strength viewpoint, that there shouldn't be much difficulty distinguishing between bridging and non-bridging oxide ions. A bridging oxide ion is simply an $O^{2-}$ with two strongly binding cations already attached to it, and thus does not want to bind anything else very strongly. The main message from Figure 4, however, is that the bridging oxygen atoms are much less basic than would be indicated in the orthosilicate-chain silicate-sheet silicate trend established in $M_4SiO_4$, $M_2SiO_3$, and $M_2Si_2O_5$ compositions. Once the possibility of binding to *any* non-bridging oxygen is lost, the basicity of the silicate is drastically reduced, much more than would be expected based on simple Pauling bond strength correlations. In this sense, the bridging oxide ion behaves almost like a new chemical species, as emphasized in Ref. 11. This issue is reflected in the failure of the pair potential model to even qualitatively reproduce the reactivity trends.

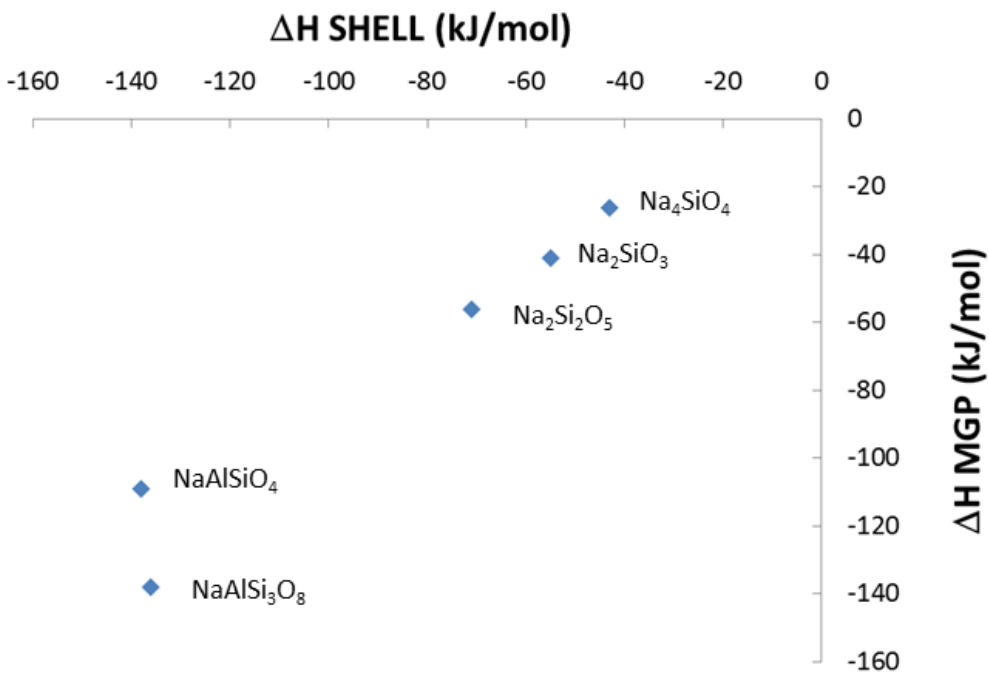

**Figure 3.** Correlation of the enthalpy of the alkali exchange reactions as calculated from the shell model potentials versus the enthalpy calculated from the Materials Genome Project database.

**Table 2.** Correlation of MGP-calculated exchange enthalpies against different measures of melt basicity: optical basicity and average Pauling bond strength.

|  | MGP (kJ/mol of alkali) | Optical basicity[a] | Avg. PBS[b] |
|---|---|---|---|
| $Na_4SiO_4$ | -26 | 0.815 | 1 |
| $Na_2SiO_3$ | -41 | 0.703 | 1.5 |
| $Na_2Si_2O_5$ | -56 | 0.614 | 1.75 |
| $NaAlSiO_4$ | -109 | 0.609 | 1.75 |
| $NaAlSi_3O_8$ | -138 | 0.554 | 1.86 |

[a]Duffy and Ingram (2002)

What physics does it take to get the right answer? Duffy and Ingram[11] focus on the changes in the polarizability and/or effective charge of the oxide ion as a function of its environment. The idea behind the optical basicity concept, that the *polarizability* (i.e. not just the state of polarization) is a strong function of the environment, is a rather sophisticated one that hasn't been implemented, at least in any consistent manner, in parameterized force fields (any force field claiming to do this should be able to reproduce, at least qualitatively, Figure 3 in Ref. 12). For a bridging oxygen atom, one imagines that the two silicon atoms are effectively sucking charge from the oxide ion, giving it a less negative overall charge, and strongly reducing the basicity. To recover such effects, one would have to use electronegativity equalization or charge equilibration–type potentials[12,13], and relate the charge to the polarizability in some manner. It's not surprising then that both the pair potential model and the Pauling bond strength correlations fail. As previously shown in Figure 3, the shell-model potentials succeed quite well in describing reactivity across the orthosilicate-tectosilicate series. In the pair potential and Pauling bond strength models the oxide ion binds poorly because it has two nearby highly charged silicon atoms. In the polarizable oxide ion model, the situation is the same, except that in this case the oxide ion is also strongly polarized by the two bound silicon ions, with the positive end of the dipole pointing towards the alkali ion, making the bridging oxygen a much worse base than in the Pauling model or the pair potential model. Even though no charge is transferred out of the oxygen atom in the shell-model potential, it still ends up being a sufficiently poor base, not because of the removal of charge, but because of the induced dipole moment. The combined charge transfer and polarization effects become very hard to untangle in any quantitative way, except to say that both effects are, of course, included, in principle, in the density functional calculations. These ideas are summarized in Figure 5.

**Implications**

The reactivity trends in the series of silicate minerals presented here allow unambiguous evaluation of different modeling approaches which could be used to estimate the basicity of silicate melts as revealed through the Na versus K binding preferences and can be used as a useful guide to future studies. In the absence of the information presented here, one might, for example, devise a program to simulate a series of glass compositions forming an interface with $KNO_3$ with a pair-potential model and calculate the extent of exchange in some large molecular dynamics calculation, perhaps with sophisticated free energy estimation methods. One could then rank-order the glasses in terms of their Na/K affinity and use this as a guide to investigating new glass compositions. The inability of the pair potential model to even rank-order the alkali exchange energies for the crystalline Na-K silicate phases means that such a program of study would be a waste of time. Some exchange would certainly be observed, due to the mixing entropy, but it would simply be an expensive way to compute $\Sigma_i X_i \ln X_i$ because, as demonstrated here, the pair potentials are not up to the task of even qualitatively describing melt basicity.  One would have to use a polarizable potential, but even this would have to have its parameters adjusted to recover the difference in basicity between $NaAlSiO_4$ and $NaAlSi_3O_8$.

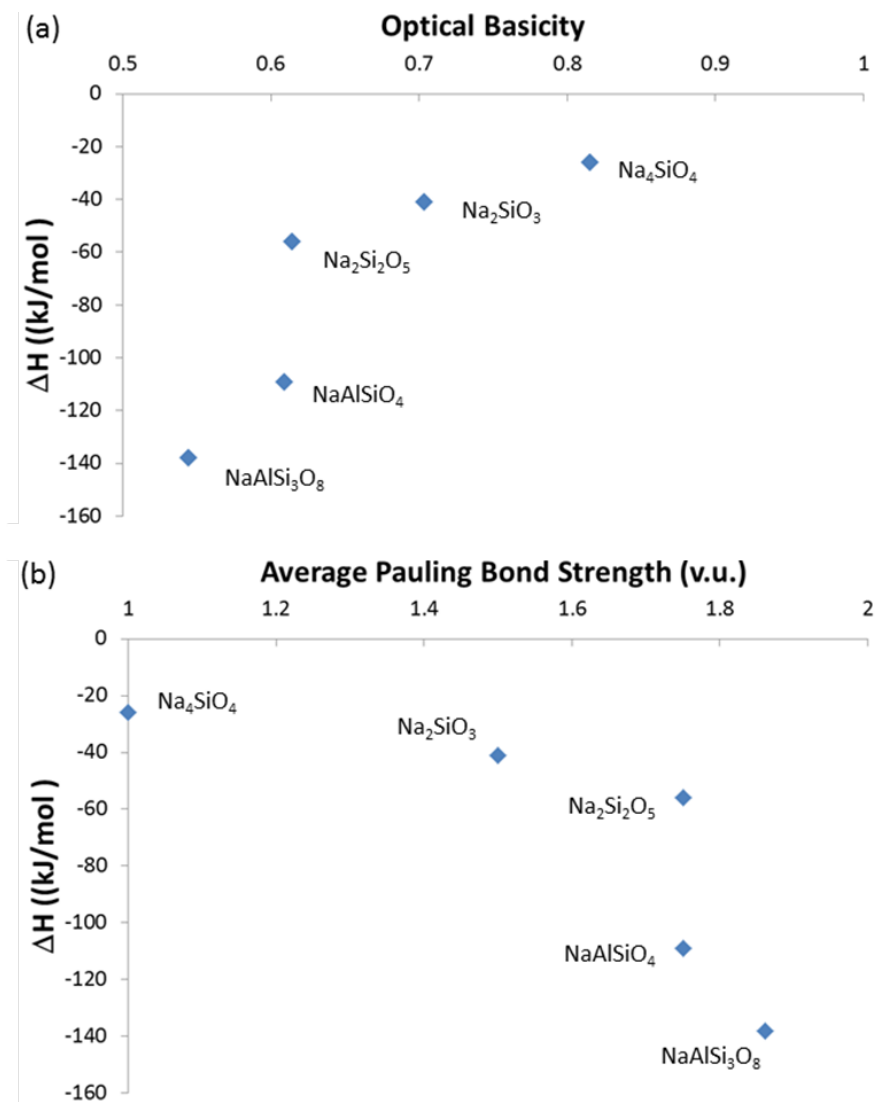

**Figure 4**. Correlation of (a) optical basicity and (b) average Pauling bond strength with the enthalpy of the alkali exchange reaction taken from the Materials Genome Project.

In the end, one would probably be better off just using average Pauling bond strength or optical basicity to make such estimates, while recognizing that these approaches cannot quantitatively describe the complexity of the oxide ligand across its diverse range of bonding environments.  At least one can rank-order the crystalline compounds correctly, and the calculations take much less time.

The first-principles approaches are very promising but they can only treat systems of 100-200 atoms and even with this limitation require significant computational resources. Even for the DFT calculations, the problem of the discrepancy between theoretical and experimental reaction energies as exhibited in Table 1 needs to be better understood. One can get around this discrepancy to some extent by settling for a correlation between theory and experiment, but with only four points in this correlation, it is hard to assess how generally reliable it would be in real design applications. Correlative relationships have proven useful, for example, in predicting the enthalpies of formation of the rare earth phosphates[14]. A more comprehensive understanding of the systematic errors inherent in the MGP electronic structure calculations will emerge as the database grows and correlations become more sophisticated.

## Summary

The following points have been demonstrated:

(1) First-principles density functional calculations do an excellent job of correlating the differences in basicity in going from compositions having four, two, one, and zero non-bridging oxygen atoms.
(2) For atomistic calculations with empirically parameterized potential functions, the non-polarizable pair potentials completely fail to account for the difference in basicity between bridging and non-bridging oxygen atoms. The shell-model potentials do much better than the pair potentials, but still fail on key test cases, like the difference in basicity between Si-O-Si and Al-O-Si oxygen atoms.
(3) Empirical correlation methods using average Pauling bond strength or optical basicity fail to account for the large difference in basicity between bridging and non-bridging oxygen atoms, but are able, at least, to correctly rank order the minerals from most basic to least basic.

The calculations presented here are easy to reproduce and should serve as a convenient benchmark for molecular modeling of ion exchange reactions in silicate glasses.

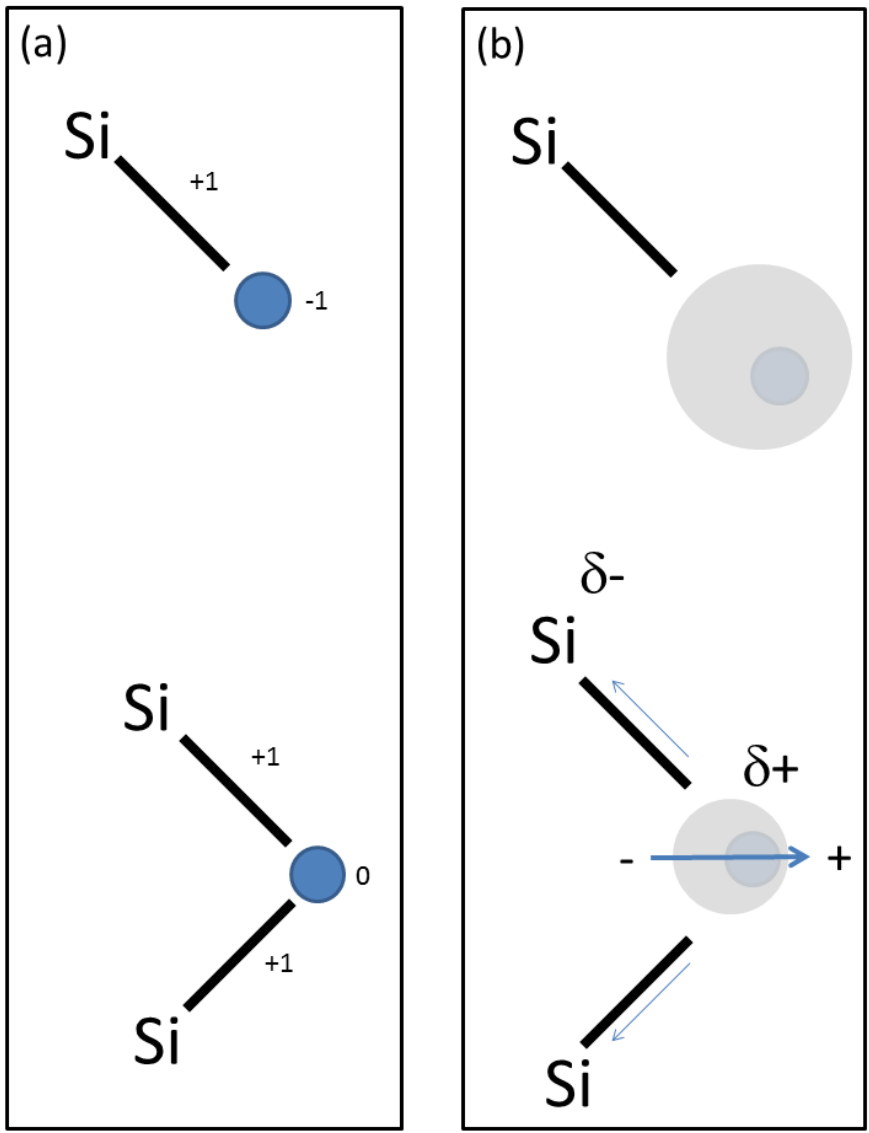

**Figure 5.** Models for differences in basicity of bridging versus non-bridging oxide ions (a) Pauling bond strength/pair potential model (b) charge-transfer (as exhibited in the first principles calculations) and polarizable oxide ion model